\documentclass[prl,twocolumn,preprintnumbers,amsmath,amssymb]{revtex4}

\usepackage{graphicx}
\usepackage{dcolumn}
\usepackage{bm}

\begin{document}

\title{Strongly Correlated Superconductivity}

\author
{M. Capone$^{1}$, M. Fabrizio$^{2,3}$, C. Castellani$^1$, E. Tosatti$^{2,3}$}
\date{\today}
\affiliation{$^{1}$ 
Dipartimento di Fisica, Universit\`a di Roma ``La Sapienza'',
and  INFM Center for Statistical Mechanics and Complexity,
Piazzale Aldo Moro 2, I-00185 Roma, Italy}

\affiliation{$^{2}$  International School for Advanced Studies (SISSA-ISAS), and INFM, Trieste-SISSA, Via Beirut 2-4, I-34014, Trieste, Italy}

\affiliation{$^3$ International Center for Theoretical Physics (ICTP),
P.O. Box 586,  I-34014 Trieste, Italy}



\begin{abstract}
High temperature superconductivity in doped Mott insulators such as the
cuprates contradicts the conventional wisdom that electron repulsion
is detrimental to superconductivity. Because  doped fullerene conductors 
are also strongly correlated, the recent discovery of 
high-critical-temperature, presumably $s$-wave, 
superconductivity in C$_{60}$ field effect devices is even more puzzling.
We examine a dynamical-mean-field solution of a model for electron 
doped fullerenes which shows how strong correlations can indeed 
enhance superconductivity close to the Mott transition. 
We argue that the mechanism responsible for this enhancement could be 
common to a wider class of strongly correlated models, including those 
for cuprate superconductors.     
\end{abstract}

\maketitle
 
In conventional superconductors, the repulsive Coulomb interaction between  
electrons tends to oppose to phonon-mediated pairing, so that the actual 
critical temperature of the onset of superconductivity ($T_c$)
decreases upon increasing electronic 
correlations. On the contrary, in the high-$T_c$ superconductors, 
strong electron-electron correlations do 
not suppress superconductivity, but rather seem to favor it because they 
are mostly poised on the brink of a repulsion-driven 
metal- Mott insulator transition (MIT). 
A recently developed approach capable of describing this transition
is the so-called Dynamical Mean-Field Theory (DMFT), where 
spatial fluctuations are neglected, but the time-dependent
quantum fluctuations are fully described\cite{GeorgesKotliar}.
As shown by DMFT,  close to a Mott transition the 
large Coulomb repulsion $U$ causes the effective metallic bandwidth  
$W$ to be dramatically renormalized to a quasiparticle  
bandwidth $W_*=ZW$, where $Z \ll 1$ is the quasiparticle residue.  
This small effective bandwidth corresponds to an increased  
quasiparticle density of states at the Fermi level $\rho=\rho_0/Z$,  
which could at first sight be thought to enhance the attractive coupling 
$\lambda = \rho V$ and thus the critical temperature 
($\rho_0$ is the bare density of states at the Fermi energy per spin, 
$V$ the pairing attraction). 
However, a decreasing $Z$ does not automatically    
turn into an increase of $\lambda$, because the pairing attraction 
$V$ is itself renormalized down, by a factor $Z^2$ within  
Migdal-Eliashberg theory,  
so that the increase of $U$ finally depresses $T_c$, an effect
further reinforced by a rising Coulomb pseudo-potential $\mu_*$. 
 
An enhancement instead of a decrease of $T_c$ with increasing Coulomb 
repulsion could 
take place if, on the contrary, the quasiparticle attraction did not  
get renormalized by $Z$, and the repulsion $U$ instead did. This  
is no doubt an appealing scenario, alas one which is at odds with all  
na\"{\i}ve expectations based on Landau Fermi-liquid theory. 
In this work we show that this scenario is actually 
viable. By solving a model for electron-doped C$_{60}$, and closely  
examining the superconductivity arising there in proximity of 
the Mott transition, we find that correlations can indeed lead  
to a huge enhancement of phonon-driven superconductivity with  
respect to the uncorrelated case. The analogy in the physics 
and even in the phase diagram of this fullerene model as a function  
of decreasing bandwidth with that of cuprates for decreasing doping  
draws a conceptual link between the two systems.   
 
The Hamiltonian describing this system is 
\begin{equation} 
H =  
\sum_{RR'}\sum_{i,j=1}^3\sum_\sigma  
t_{RR'}^{ij} c^\dagger_{R,i\sigma}c^{\phantom{\dagger}}_{R',j\sigma} 
+ {\frac{U}{2}}\sum_R n_R n_R + H_{Hund}, 
\label{Ham} 
\end{equation} 
where $c_{R,i\sigma}$ is the electron annihilation operator at site $R$  
in orbital $i$ ($i=1,2,3$) (the $t_{1u}$ level in  
C$_{60}$ is three-fold degenerate) 
with spin $\sigma$, and  $n_R=\sum_{i,\sigma} n_{R,i\sigma}$, where
$n_{R,i\sigma}=c^\dagger_{R,i\sigma}c^{\phantom{\dagger}}_{R,i\sigma}$ 
is the electron occupation number. We also assumed for simplicity  
$t_{RR'}^{ij}=\delta_{ij}t_{RR'}$.  
We introduce the angular momentum 
density operators $L_{i,R} = \sum_{j,k,\sigma} c^\dagger_{R,j\sigma} 
\hat{L}_{i,jk}c^{\phantom{\dagger}}_{R,k\sigma}$, with  
$\hat{L}_{i,jk}=-i\varepsilon_{ijk}$ proportional to the Levi-Civita tensor,  
and the spin density operators  
$S_{i,R} = 1/2\sum_{k,\alpha,\beta} c^\dagger_{R,k\alpha} 
\hat{\sigma}_{i,\alpha\beta} c^{\phantom{\dagger}}_{R,k\beta}$,  
with $\hat{\sigma}_i$ ($i=1,2,3$) the Pauli matrices. 
In terms of these operators, Hund's term is  
$H_{Hund} = -J_H\sum_R \left(2\vec{S}_R\cdot\vec{S}_R  
+ \frac{1}{2}\vec{L}_R\cdot\vec{L}_R\right) + \frac{5}{6} 
\left(n_R - 3\right)^2$. 
 
The bare $J_H$ is positive. However, in fullerene, the 
Jahn-Teller coupling of electrons, and presumably also of holes 
\cite{Manini01}, to the $H_g$ molecular vibrations can reverse 
Hund's rules, favoring low spin and angular momentum ground states. 
We include this crucial electron-phonon effect by assuming $J_H<0$, 
formally equivalent to treating the Jahn-Teller coupling in 
the antiadiabatic limit, where it can be shown to renormalize 
$J_H \to J_H - 3E_{JT}/4<0$, with $E_{JT}$ the Jahn-Teller energy 
gain. The antiadiabatic approximation is justified for 
fullerene where vibron frequencies are as high as 0.1 eV, to be 
compared with a correlation-narrowed quasiparticle bandwidth $ZW$, 
where the bare bandwidth $W \sim$ 0.5 eV and a quasiparticle  
residue $Z \ll 1$, due to a very large $U/W$. In any case, the  
neglect of retardation disfavors superconductivity, by preventing 
high-energy screening of the repulsion, hence overestimating 
$\mu_*$. 
 
Following Ref. 3 we studied model (\ref{Ham})  
by DMFT\cite{GeorgesKotliar}, varying $U/W$,  
at a fixed ratio $J_H/U = -0.02$ and integer filling $\langle n\rangle=2$
(or, equivalently, $\langle n \rangle =4$). 
At weak coupling, $U\ll W$, model (\ref{Ham}) describes a metal with  
three 1/3-filled degenerate bands. If alone, the negative $J_H$ 
would develop a superconducting $s$-wave order parameter 
$\Delta_R = \sum_{i=1}^3 c^\dagger_{R,i\uparrow} 
c^\dagger_{R,i\downarrow}$. With $J_H=0$ and considering explicitly  the 
electron-phonon coupling, Migdal-Eliashberg and DMFT calculations at
relatively small $U/W$ have well characterized  
this conventional, weakly or moderately correlated superconducting  
phase\cite{GunnarssonRMP,Han,Han2}. 
However, for our present $J_H/U = -0.02$ the  
effective superconducting coupling $\lambda= 10\rho_0|J_H|/3 
=0.2 \rho_0 U/3$ ($\rho_0$ is here the density of states per spin and band) 
is much smaller than the  
Coulomb pseudopotential $\mu_* = \rho_0 U$, and weak coupling 
superconductivity is suppressed in favor of a normal metal.  
At strong coupling, $U\gg W$, the system is a Mott insulator.  
Each site is occupied by two electrons which, since $J_H <0$, 
form  a spin and orbital singlet, as expected in a Mott-Jahn-Teller 
insulator\cite{Fabrizio97}). This state, a kind of on-site version of the 
Resonant Valence Bond (RVB) state\cite{Anderson}, 
is nonmagnetic and has a gap to spin,  
orbital, and charge excitations. The transition between the  
metal and the strong coupling Mott insulator is however not direct, and  
a superconducting phase is known to intrude in between 
\cite{Capone01}. The properties of this superconducting phase  
are, we now find, striking. 
 
Fig. 1 shows the superconducting gap $\Delta$, obtained 
as the zero-frequency anomalous self-energy, compared 
with the hypothetical superconducting  gap calculated in
standard Bardeen-Cooper-Schrieffer (BCS) theory at $U=0$, and with the 
actual spin gap $\Delta_{spin}$ extracted as the edge of the main spectral peak
in the dynamical spin susceptibility. 
 
\begin{figure}[htbp]
\begin{center}
\includegraphics[width=8cm]{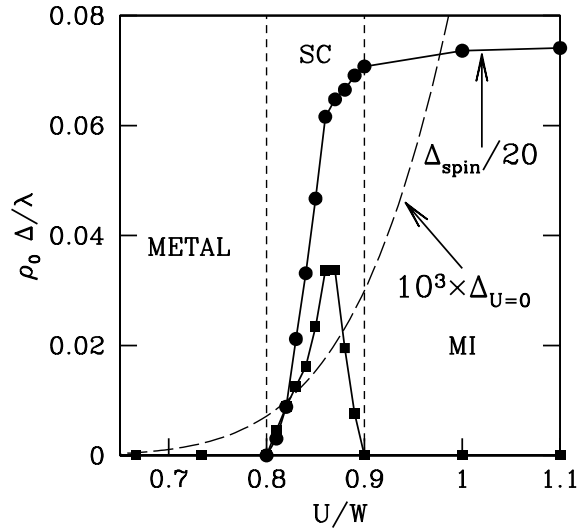}
\end{center}
\caption{Superconducting gap $\Delta$ (squares) as a function 
of $U/W\equiv 50 |J_H|/W$.  
SC and MI stand for superconductor and Mott insulator, respectively.  
Also shown are the spin gap $\Delta_{spin}$, reduced by a factor 20 
(circles),  and the BCS gap calculated as  
function of $50 |J_H|/W$ at $U=0$ and multiplied by 1000 (long-dashed line).
Gaps are normalized to $\lambda/\rho_0 = 10 \vert J_H \vert/3$, 
which measures the pair attraction.}
\label{fig1}
\end{figure}

Superconductivity is seen to arise suddenly out of the normal metal 
upon increasing repulsion above a critical value $(U/W)_c$ (here 0.8), 
and below the MIT (here at 0.9). At $(U/W)_c$ the  
superconducting gap initially coincides with the spin gap, as 
in weak coupling  BCS theory, but the two rapidly deviate. The 
larger spin gap merges with that of the Mott state; $\Delta$ instead 
reaches a peak value -- a huge 1000 times the $U=0$  BCS gap 
calculated for the same pairing attraction -- before falling again 
down to zero at the MIT. The large peak value of $\Delta$ is of 
the order of the maximum value which could be reached at $U=0$ if 
the bandwidth were comparable with $|J_H|$, which is also the condition 
to get the maximum $T_c$ for a fixed unretarded attraction
\cite{Robaskiewicz}. In terms of a Landau Fermi-liquid description 
of the metallic phase, this suggests that the quasiparticles close 
to the MIT have an effective bandwidth $W_* \sim ZW\sim |J_H|$, 
and experience an attraction of the very same order of magnitude. 
If this were indeed the case, the main effect of strong correlations 
would be to decrease $W_*$ and thus increase the quasiparticle mass, 
leaving behind only a small residual quasiparticle repulsion. If 
the attractive vertex remained at the same time substantially  
unrenormalized while $Z \to 0$, then the quasiparticle scattering  
amplitude would switch from repulsive at weak coupling to attractive  
close to the MIT. A lack of renormalization of $J_H$ is plausible,  
because Hund's coupling does not compete with $U$ but rather  
benefits from it. In fact, $U$ brings the system toward  
the atomic limit where Hund's rules are obeyed, whereas  
the metallic phase is where they are violated. There is  
here a similarity to the $t-J$ model of cuprates, where $J$ is 
also apparently unrenormalized close to the insulator, as 
suggested by slave boson methods \cite{Kotliarslave} and 
by numerical calculations \cite{sorella}.
 
We found that this appealing, but thus far hypothetical scenario,  
is confirmed by a careful analysis of the metallic phase within  
Landau Fermi-liquid theory. DMFT enables a study of the normal 
metal even inside the superconducting region, by preventing 
spontaneous breaking of the gauge symmetry in the self-consistency 
equations, and providing a full description of the quasiparticles 
and of their mutual interactions close to the Mott transition. 
The Landau functional of the model, which possesses spin SU(2) and 
orbital O(3) symmetry, contains here 
a multiplicity of Landau parameters $f^{S(A)}$, $g^{S(A)}$ and  
$h^{S(A)}$\cite{Capone01}.   
Defining $F$-parameters 
$F^{S(A)}= 6\rho_0 f^{S(A)}/Z$, $G^{S(A)}= 12\rho_0 g^{S(A)}/Z$ and 
$H^{S(A)}= 4\rho_0 h^{S(A)}/Z$, dimensionless quantities 
which measure the strength of
the interactions between quasiparticles, the susceptibilities have 
the standard expression 
$\frac{\chi}{\chi^{(0)}} = \frac{1}{Z}\frac{1}{1+F}$, 
where $\chi$ refers to the charge(spin) susceptibility for 
$F=F^{S(A)}$, and analogously for all the other orbital and spin-orbital 
($G^{S(A)}$ and $H^{S(A)}$ parameters)  
susceptibilities. By calculating in DMFT 
the quasiparticle residue, $Z$, and all six susceptibilities, 
we obtain the $F$-parameters of the model as a function of $U/W$.

Fig. 2 shows the decrease of $Z$ in the metallic solution  
on approaching the MIT.  
Superconductivity sets in at $Z=Z_{crit}\simeq 0.06$,  
a very small value indeed.
The charge compressibility decreases as a function of $U/W$, 
and vanishes at the MIT, consistent 
with the approaching incompressible insulator. 
The spin and all four orbital and spin-orbital susceptibilities, 
which initially increase at small $U/W$ (Stoner enhancement), 
turn around at $U/W \sim 0.7$, eventually vanishing at the MIT, consistent 
with a spin and orbital gap in the insulator. 
Accordingly, $F^S$ monotonically increases from $\sim U/W$ at  
weak coupling to infinity at the MIT, while the other parameters  
$F^A$, $G^{S(A)}$ and $H^{S(A)}$ start off 
negative proportional to $-U/W$  
roughly until $U \sim ZW$, but then
turns upward, cross zero, and finally diverge 
like $1/Z^2$ at the MIT. 
 
\begin{figure}[htbp]
\begin{center}
\includegraphics[width=8cm]{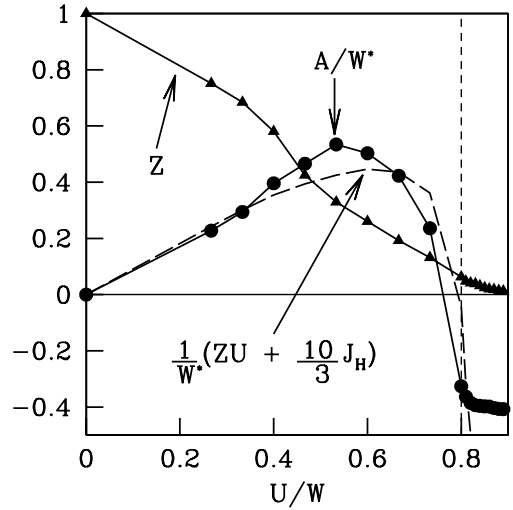}
\end{center}
\caption{Quasiparticle residue $Z$ (triangles) as function of 
$U/W$. The vertical dashed line at $U/W=0.8$ identifies the critical value 
above which the  metallic solution may spontaneously develop a 
superconducting order parameter. 
The quasiparticle scattering amplitude $A$ (circles) and its heuristic 
approximation $ZU+10J_H/3$ (long-dashed line) are also shown, both 
crossing zero at the metal-superconductor transition.}
\label{fig2}
\end{figure}

This behavior of the Landau parameters in the metallic phase is at 
the root of the superconducting instability, as is seen by calculating 
the quasiparticle pair $s$-wave scattering amplitude $A$.  
$A$ has two contributions,  $A_{i\to i}$ and $A_{i\to j}$, describing 
singlet pair scattering from orbital $i$ into 
the same or into another orbital, respectively. They are given by 
$A_{i\to i} = 
\frac{Z}{12\rho_0} 
\left( \frac{F^S}{1+F^S} -3\frac{F^A}{1+F^A} 
+2\frac{G^S}{1+G^S} -6\frac{G^A}{1+G^A}\right)$, 
and $A_{i\to j} = 
\frac{Z}{8\rho_0} 
\left( -\frac{H^S}{1+H^S} +3\frac{H^A}{1+H^A} 
+\frac{G^S}{1+G^S} -3\frac{G^A}{1+G^A}\right)$. 
Fig. 2 shows the pair amplitude 
$A=A_{i\to i}+2A_{i\to j}$. At weak coupling, $A_{i\to i}=U+4J_H/3>0$ and 
$A_{i\to j}=J_H$, so that $A = U+10J_H/3$ is repulsive. 
However, as the MIT is approached and all $F$-parameters diverge, 
$A\to -Z/2\rho_0$, attractive and about equal to half of the quasiparticle 
bandwidth $W_*/2=ZW/2$, confirming our proposed scenario. 
In fact, the assumption of a quasiparticle repulsion  
renormalized by $Z$\cite{ansatz}, and an unrenormalized attraction 
$J_H$ would imply here $A\simeq ZU+10J_H/3 $.  
This simple expression is seen to compare remarkably well with  
the true $A$ up to $(U/W)_c$  leading to a very accurate  
estimate of 0.067 for $Z_{crit}$. 
 
The crossover from weak to strong correlations occurs  
when the lower and upper Hubbard bands separate from each  
other, uncovering the quasiparticle resonance 
in the spectral function. This suggests 
a two-component description of the model, similar to that 
used to analyze the MIT in terms of Kondo effect\cite{Gabi}. 
We find that a two-component model describes very well the  
strongly correlated superconducting phase, in particular  
the probability $P(n)$ for a site to possess $n$ electrons 
in the ground state. We calculated $P(n)$ -- the total
weight of states with occupancy $n$ in the ground state --  
for the superconductor (Fig. 3 (B),(C),(D)), 
and that for the nearby Mott insulator 
$P_{ins}(n)$ (Fig. 3 (A)) to find that they are  quite similar.  
In spite of an exceedingly large $\Delta$, there is no evidence of  
preformed pairs or bipolarons in the superconductor, as underlined by  
the strong steady peaking of $P(n)$ around $n=2$.

By assuming the ``two-component'' form  
$P(n) = Z P^{(SC)}_{qp}(n) + (1-Z)P_{ins}(n)$, where $Z$, $P(n)$ and
$P_{ins}(n)$ are known, we extracted 
the quasiparticle probability distribution 
$P^{(SC)}_{qp}$ in the strongly correlated superconductor
between $U/W=0.8$ and $U/W=0.9$ (Fig. 3 (A$^\prime$)(B$^\prime$)(C$^\prime$)(D$^\prime$)). 
It shows strong oscillations between even and odd $n$,  
as expected for a superconductor, with no major 
variations as a function of $U/W$ even close to the Mott transition
($U/W=0.9$), 
consistent with a weak to intermediate coupling superconductivity,  
implied by $A\simeq -W_*$. 
As a check, $P^{(METAL)}_{qp}(n)$ in the (metastable) non superconducting 
metal is also extracted (Fig. 3 (A$^{\prime\prime}$)(B$^{\prime\prime}$)(C$^{\prime\prime}$))
and found to be similar to that of a free 
Fermi liquid (Fig. 3 (D$^{\prime\prime}$)), 
indicating almost free quasiparticles. The accuracy shown 
by this check is remarkable, as 
$P^{(METAL)}_{qp}$ is but a tiny fraction $\sim Z$ of $P(n)$ and  
there is no free parameter.  
We conclude that, in the strongly correlated superconductor, 
free-fermion-like quasiparticles of weight $Z$ become strongly 
paired while floating in a prevailing Mott insulator background. 
That background slows them down while taking away their Coulomb 
repulsion, but not their on-site (here Jahn-Teller originated) 
pair attraction.  
Somewhat similar to systems with spin-charge separation,
the charge degrees of freedom are strongly renormalized close
to the Mott transition, but the spin degrees of freedom --
here including the pairing attraction -- are not.
The phase diagram of Fig. 1 for increasing $U$ bears 
a remarkable similarity to that of cuprates for decreasing doping.
We believe the superconductivity in the $t-J$ model of cuprates 
to be in fact of a deeply similar origin -- although the intersite
antiferromagnetic interaction does of course introduce important  
differences over our on-site $J_H$. The mechanism inducing singlet 
formation without competition with the Coulomb repulsion $U$ is 
not far in spirit from Anderson's original RVB idea for cuprates
\cite{Anderson}.

Coming to fullerenes, we are only beginning to explore the full phase 
diagram and calculate $T_c$ and other properties for variable  
electron, and also hole doping. The possibility that superconductivity 
in these systems could be of the present, strongly correlated kind, 
seems real. 
Even if our solution is obtained for $\langle n\rangle = 2$ or $4$
(where superconductivity has not yet been found), while
the investigation of the $\langle n\rangle = 3$ case (where superconductivity
is actually observed) requires further work,
we expect a very similar scenario also for the latter case.
In particular, the chemically expanded $\langle n\rangle = 3$ system
(NH$_3$)K$_3$C$_{60}$ is experimentally found to be insulating
with low-spin ($S=1/2$), rather than high-spin ($S=3/2$)
as expected for a regular Mott state\cite{tou}.
The low-spin is clearly of Jahn-Teller origin,
indicating a Mott-Jahn-Teller insulator,
exactly as in the $\langle n\rangle = 4$ case of K$_4$C$_{60}$.

\begin{figure}[htbp]
\begin{center}
\includegraphics[width=8cm]{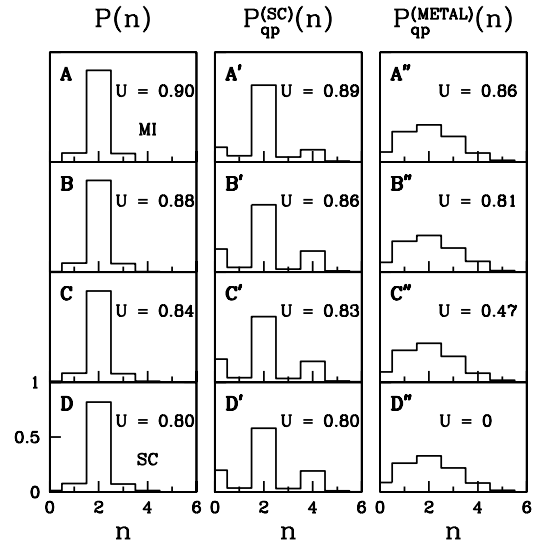}
\end{center}
\caption{Occupation probabilities for the particles, $P(n)$, and for 
the quasiparticles in the superconducting   
and in the metallic phases, $P_{qp}^{(SC)}(n)$ and $P_{qp}^{(METAL)}(n)$, 
respectively.}
\label{fig3}
\end{figure}

An important detail in fullerenes is the actual value of the 
superconducting $\lambda$. Strict electron-phonon coupling would yield
a realistic value of $\lambda\sim 0.8-1.1$\cite{GunnarssonRMP}.
If alone, this large coupling would place fullerene superconductors 
in the intermediate coupling regime already at $U=0$. There, a further
increase of $\lambda$ should not really raise much $T_c$, or
might even reduce it, in contrast with the well-known strong
increase of $T_c$ with increasing volume\cite{yildirim}.
An independent estimate of the effective pair attraction can
be obtained by comparing the spin gap observed both in insulating 
K$_4$C$_{60}$ \cite{mehring} and in superconducting K$_3$C$_{60}$ 
\cite{brouet} $\Delta_{spin} \simeq
0.07-0.1$ eV with that of our $J_H < 0$ model, through 
$\Delta_{spin} \simeq 5|J_H|$\cite{Capone00}.
We get in this way $J_H \simeq -0.02~{\rm eV} \simeq -0.02 U$, 
hence $\lambda\simeq 0.13$, which is the tentative value adopted here. 
This large reduction of the effective $\lambda$ must, as explained above,  
be due to strong cancellation by the bare Hund's rule $J_H \sim 0.05 eV$
\cite{martin}. With a reduced $\lambda$ and large $U$, the
only way to explain superconductivity and its increasing $T_c$ with 
volume (increasing $U/W$), is to invoke the strongly correlated 
superconductivity we found just above $(U/W)_c$. 
For further volume expansion, our
model predicts as in Fig. 1 an eventual decline of $T_c$, 
and a Mott insulator for integer filling. Both features are
observed in ammoniated
compounds of the K$_3$C$_{60}$ family\cite{NH3C60}. In the K$_4$C$_{60}$
family, conversely, $U/W$ is above the MIT value, and we have 
Mott-Jahn-Teller insulators. 
Finally, we surmise that a similar strongly correlated superconductivity, 
 modified to account for the $d=5$ degeneracy of the 
$h_u$ hole states, should be relevant to the recently discovered $C_{60}(n+)$
superconductivity\cite{Batlogg}. Here a somewhat stronger
electron-phonon coupling has been observed\cite{Batlogg} and calculated
\cite{Manini01}, whereas the hole bandwidth and intra-molecular Coulomb
repulsion are most likely similar to those of electrons. 
The enhanced superconductivity in C$_{60}$CHCl$_3$ and C$_{60}$CHBr$_3$
expanded lattices could result from the increase of $U/W$ and 
of electronic correlations, whereas the alternative explanation
of  a BCS-like increase in
the density of states has been put in deep question by very 
recent results\cite{dinnebier}.
The full 
development of the theory of C$_{60}(n+)$ superconductivity 
is a task we reserve for 
the near future.\cite{acknowl}


\begin{thebibliography}{999}
\bibitem{ansatz} This assumption is inspired by the evidence  
that the quasiparticle contribution to the average value of the double 
occupancy in the single band Hubbard model scales approximately 
like $Z$ close to the MIT\cite{GeorgesKotliar}.
\end{thebibliography}
\end{document}